\documentclass{jaa}
\usepackage{natbib}
\usepackage{graphicx}

\def\pasp{PASP} 
\def\apj{ApJ} \def\mnras{MNRAS}
  \def\aj{aj} 
 \def\apjs{ApJS}  
\def\nat{Nat}

\begin{document}\sloppy

\title{Short-Timescale Variability of the Blazar Mrk 421 from AstroSat and Simultaneous Multi-Wavelength Observations}


\author{Ritaban Chatterjee\textsuperscript{1,*}, Susmita Das\textsuperscript{1}, Archishman Khasnovis\textsuperscript{1}, Ritesh Ghosh\textsuperscript{2}, Neeraj Kumari\textsuperscript{3,4}, Sachindra Naik\textsuperscript{3}, V.~M.~Larionov\textsuperscript{5,6}, T.~S.~Grishina\textsuperscript{5}, E.~N.~Kopatskaya\textsuperscript{5}, E.~G.~Larionova\textsuperscript{5}, A.~A.~Nikiforova\textsuperscript{5,6}, D.~A.~Morozov\textsuperscript{5}, S.~S.~Savchenko\textsuperscript{5}, Yu.~V.~Troitskaya\textsuperscript{5}, I.~S.~Troitsky\textsuperscript{5}, and A.~A.~Vasilyev\textsuperscript{5}}
\affilOne{\textsuperscript{1}Department of Physics,  Presidency University, 86/1 College Street, Kolkata 700073.\\}
\affilTwo{\textsuperscript{2}Visva-Bharati University, Santiniketan, Bolpur, West Bengal, India 731235. \\}
\affilThree{\textsuperscript{3}Astronomy and Astrophysics Division, Physical Research Laboratory, Navrangapura, Ahmedabad 380009, Gujarat, India. \\}
\affilFour{\textsuperscript{4}Department of physics, Indian Institute of Technology, Gandhinagar 382355, Gujarat, India. \\}
\affilFive{\textsuperscript{5}Astronomical Institute of St. Petersburg State University, Universitetskij Prospekt 28, Petrodvorets, 198504 St. Petersburg, Russia. \\}
\affilSix{\textsuperscript{6}Pulkovo Observatory, St.-Petersburg, Russia. \\}

\twocolumn[{

\maketitle

\corres{ritaban.physics@presiuniv.ac.in}

\msinfo{---}{---}

\begin{abstract}
We study the multi-wavelength variability of the blazar Mrk 421 at minutes to days timescales using simultaneous data at $\gamma$-rays from Fermi, 0.7$-$20 keV energies from AstroSat, and optical and near infrared (NIR) wavelengths from ground based observatories. We compute the shortest variability timescales at all of the above wave bands and find its value to be $\sim$1.1 ks at the hard X-ray energies and increasingly longer at soft X-rays, optical and NIR wavelengths as well as at the GeV energies. We estimate the value of the magnetic field to be 0.5 Gauss and the maximum Lorentz factor of the emitting electrons $\sim$$1.6 \times 10^5$ assuming that synchrotron radiation cooling drives the shortest variability timescale. Blazars vary at a large range of timescales often from minutes to years. These results, as obtained here from the very short end of the range of variability timescales of blazars, are a confirmation of the leptonic scenario and in particular the synchrotron origin of the X-ray emission from Mrk 421 by relativistic electrons of Lorentz factor as high as $10^5$. This particular mode of confirmation has been possible using minutes to days timescale variability data obtained from AstroSat and simultaneous multi-wavelength observations.
\end{abstract}

}]


\doinum{12.3456/s78910-011-012-3}
\artcitid{\#\#\#\#}
\volnum{000}
\year{0000}
\pgrange{1--}
\setcounter{page}{1}
\lp{1}

\section{Introduction}
Relativistic jets pointed toward the line of sight of the observer is a defining property of blazars \citep{urr95}, a class of radio loud active galactic nuclei (AGN). Due to relativistic beaming the apparent luminosity of the jet is amplified by a factor of 10 to 10$^4$ in the observer's frame.  Consequently, emission from the other parts of the AGN such as the accretion disk, broad line region, and dusty torus are overwhelmed by that from the jet. Blazars are often bright over a large range of wavelengths from radio to GeV or even TeV $\gamma$-rays. The spectral energy distribution (SED) of blazars are characteized by two broad peaks: one at the IR-X-ray wavelength range due to synchrotron radiation by relativistic electrons present in the jet \citep{bre81,urr82,imp88,mar98} and the other in the GeV range sometimes extending to TeV energies. In the so called ``leptonic scenario,'' the high energy peak may be due to inverse-Compton (IC) scattering of ``seed'' photons from the broad line region, dusty torus or the jet itself \citep{mar92,chi02,arb05,sik94,cop99,bla00,der09,bot10,rom17} by the same relativistic electrons that are generating the synchrotron radiation for the lower energy peak. Alternatively, in the so called ``hadronic model,'' X-rays and $\gamma$-rays may be produced due to synchrotron radiation by relativistic protons, which are accelerated along with the electrons in the jet, proton-induced particle cascades, or interactions of these high-energy protons with external radiation fields \citep{muc01,muc03,bot13}. Based on the location of the synchrotron peak ($\nu_{synch}$) in the broad band SED, blazars are divided into three categories: low, intermediate and high synchrotron peaked: LSP ($\nu_{synch} < 10^{14}$ Hz), ISP ($10^{14} < \nu_{synch} < 10^{15}$ Hz), and HSP ($\nu_{synch} > 10^{15}$ Hz), respectively \citep[e.g.,][]{abd10_sed,bot19}.

Blazars are characterized by fast and high amplitude variability in emission, which is due to the injection of energy and fluctuations in the density, magnetic field and other physical properties of the jet originally caused by instabilities in the accretion disk \citep[e.g.,][]{mal14}. Due to the beaming effect the timescale of variability becomes faster in the obsever's frame by a factor $\sim$10. In the leptonic model, the emission at wavelengths near the two peaks in the SED are caused by the highest energy part of the electron spectrum. Therefore, variability at those wavelengths are expected to be the most pronounced, e.g., by an order of magnitude or more over months to years timescales \citep[e.g.,][]{maj19}. Blazar variability is red noise in nature, i.e., amplitude of variability is smaller at shorter timescales \citep[e.g.,][]{cha12}. But near the SED peak wavelengths, emission may vary by a factor of a few even at sub-day timescales.  

Blazar variability at X-ray and other wave bands at longer timescales have been studied extensively \citep[e.g.,][]{cha08,abd10_timing,cha12,raj20} but that at sub-day timescales has been less common. It is mainly because even the brighter blazars are not detected with high signal to noise at short exposures. Long and quasi-continuous observations with X-ray telescopes are needed to obtain comprehensive information about their short-term fluctuations. The blazar Mrk 421, one of the brightest blazars in the X-ray sky, is an ideal source for such study. It is an HSP blazar with the low energy peak of the SED at the X-ray band \citep[e.g.,][]{abd11,ban19}. Hence, the X-ray variability is expected to be very pronounced. However, averaged over the last 10 years its brightness is 10 cts/s in Swift-XRT\footnote{https://www.swift.psu.edu/monitoring/}. Hence, to probe its short-timescale properties at minutes to days timescale long stares are necessary which are not possible with Swift as it has other priorities.

In this paper, we study the minutes to days timescale X-ray variability of Mrk 421 using a quasi-continuous observation by AstroSat \citep{agr06,sin14}. We compare the same with simultaneous observations at GeV, optical, and near infrared (NIR) wave bands. We analyze the data to test if the multi-wavelength variability properties are consistent with the expectation of the standard model of blazar emission as described above. In particular, we check if the amplitude and timescale of variability are consistent with the leptonic model in which the NIR, optical, and X-ray emission in an HSP blazar such as Mrk 421 are generated by increasingly higher energy part of the electron distribution and the GeV emission is produced by lower energy electrons via IC processes. In {\S}2, 3, and 4 we describe the data reduction, present the results of the variability analysis, and discuss the implications of the results, respectively. 


\section{Data}
\subsection{X-ray and $\gamma$-Ray Data}
Mrk 421 was observed with AstroSat (Obs Id: A05\_204T01\_9000002856, PI: Ritaban Chatterjee ) on April 23-28, 2019. We received the data for individual orbits of the satellite (Level-1 data) from the Indian Space Science Data Center (ISSDC). We first processed the Level-1 Soft X-ray Telescope \citep[SXT;][]{sin16,sin17} data with the {\tt sxtpipeline} task, which is a part of the SXT software (AS1SXTLevel2, version 1.4b) available at the SXT POC Website\footnote{http://www.tifr.res.in/~astrosat\_sxt/sxtpipeline.html}. Level-2 cleaned event files for the individual orbits were extracted using pipeline calibration of the source events. The extraction process includes filtering out any contamination by the charged particles due to excursions of the satellite through the South Atlantic Anomaly region and occultation by the Earth. To select only X-rays and avoid counting charged particles, we selected only the events with grade $0-12$, i.e., single-quadruple events. Next, we merged the cleaned event files of all the orbits into a single event file using a {\tt Julia} based software tool developed by G. C. Dewangan in order to avoid any time-overlapping events in consecutive orbits. A circle of radius 15 arcmin was used to extract the source light curve in the energy interval $0.7-8$ keV. 
\begin{figure*}
\centering\includegraphics[height=.8\textheight]{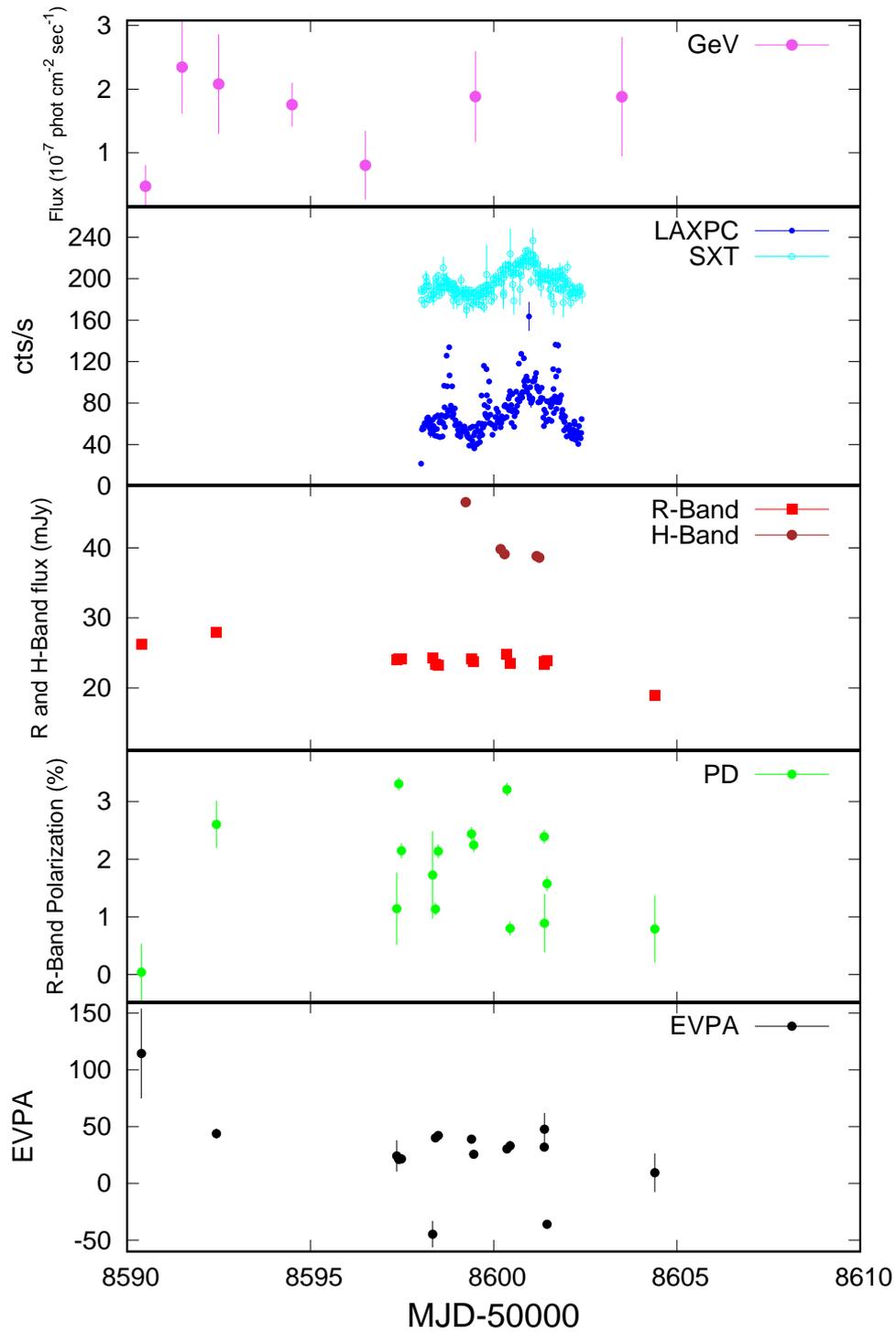} 
\caption{Variation of the 0.1$-$300 GeV $\gamma$-ray flux from Fermi-LAT in daily bins, hard (4$-$20 keV) and soft (0.7$-$8 keV) X-ray flux from AstroSat LAXPC and SXT, respectively, $H-$band flux density from Mount Abu Infrared observatory NICS instrument, and $R-$band flux density and optical polarization properties from the St. Petersburg University 40 cm LX-200 telescope and the Crimean observatory 70 cm AZT-8 telescope of Mrk 421 from 2019 April 16 - May 06. The SXT data have been scaled and shifted for clarity. The unscaled light curve is shown in the next figure. } \label{multi_all}
\end{figure*}

The Large Area X-ray Proportional Counter (LAXPC) instrument onboard AstroSat has three coaligned proportional counters, LAXPC10, LAXPC20, and LAXPC30  \citep{yad17,ant17}. We did not use the LAXPC30 data due to a gain instability issue caused by gas leakage. 
LAXPC data were collected in the Event Analysis (EA) mode. During the observation, the LAXPC10 was operating at a lower gain. Hence, we have used data only from LAXPC20 for our analysis. The light curves and spectra were generated using the LaxpcSoft software package 5. The background was estimated from blank sky observations, where there are no known X-ray sources. We extracted light curves using data from the top layer for the energy interval $4-20$ keV.
\begin{table*}[htb]
\centering
\tabularfont
\caption{Variability Timescale and Excess Variance at Different Wave Bands}\label{tau_d_exvar} 
\begin{tabular}{lcccccc}
\topline
Wave Band    & $\tau_d$ (increase) (ks) & $\Delta$t (ks)  & $\tau_d$ (decrease) (ks) & $\Delta$t (ks)   &  Normalized Excess Variance \\\midline
GeV					& 37.2$\pm$23.3					& 86.4 					& 279.1$\pm$256.6 			& 432.0 					& ---    \\
Hard X-Ray 	&	1.5$\pm$0.2						&	 2.0 						& 1.1$\pm$0.2						&	1.0 						& 0.29	 \\
Soft X-Ray 	    &	2.3$\pm$1.4						&	 0.6 						& 1.5$\pm$1.1						&	0.6 						& 0.12		\\
$R$-Band 		&	195.3$\pm$89.0				&	 6.2 						& 99.4$\pm$7.3					&	7.3 						& 0.07		\\
$H$-Band 		&	---										&	--- 						& 366.3$\pm$34.5				& 91.6 					& 0.07 \\
\hline
\end{tabular}
\end{table*}

At the GeV energies, we use data from the Large Area Telescope on board Fermi Gamma-Ray Space Telescope \citep{abd20}. Fermi has a large field of view ($\sim$2.4 Sr) and monitors the whole sky every three hours. We use the light curve at 0.1-300 GeV obtained from the Fermi Science Support Center (FSSC)\footnote{https://fermi.gsfc.nasa.gov/ssc/data/access/lat/msl\_lc/}. 


\subsection{Optical and Near Infrared Data}
The optical flux and polarimetric data were obtained using the St. Petersburg University 40 cm LX-200 telescope and the Crimean observatory 70 cm AZT-8 telescope. For the details of data analyses and processing see \citet{lar08}.

Near-infrared observations of Mrk~421 in $H$-band, simultaneous with the AstroSat observations of the source, were carried out with PRL's 1.2 m telescope at Mount Abu Infrared Observatory (MIRO), India. The photometric observations were performed on the nights of 25, 26 and 27 April 2019 in imaging mode using Near-Infrared Camera and Spectrograph (NICS) instrument. At Mt. Abu, the NICS provides $\sim$1 arcsec seeing over $8 \times 8$ arcmin$^2$ with $1024 \times 1024$ pixels array configuration. A set of five frames of $\sim$50 s were taken in 
$H$-band ($1.635-1.780$ $\mu$m, central wavelength at 1.49 $\mu$m), dithered at 5 different positions. The data were reduced using {\tt IRAF} software package as described in \citet{nai10}. In this procedure, the sky-frame was generated for a particular filter by median combining all the raw and dithered images. After subtracting skyframes from the raw images, clean images were obtained and then corrected for effects like cosmic rays. Aperature photometry was performed using the {\tt PHOT} task in {\tt IRAF}. We selected SAO 015832 and SAO 062433 as standard stars.

\section{Results}
The multi-wavelength light curves of Mrk 421 during 2019 April 16 - May 06 are shown in Fig. \ref{multi_all}. It is evident that the variability in that time interval is most pronounced in the X-ray energies. We compute a variability timescale during both increase and decrease of flux using the following formula \citep{sai13,roy19}:
\begin{equation}
\tau_{d} = \Delta t \frac{ln2}{ln(f_2/f_1)},
\end{equation}
where $f_1$ and $f_2$ are the flux values at time $t_1$ and $t_2$, and $\Delta t$ is the difference between $t_1$ and $t_2$. We compute $\tau_{d}$ for all pairs of data points in the light curves and search for the shortest value in each. The shortest timescale of variability that we find at each band along with the corresponding value of $\Delta t$ are given in Table \ref{tau_d_exvar}. Value of $\tau_{d}$ during increase of flux for $H$-band could not be obtained because the length and sampling of the light curve are not enough for that calculation. In addition, Table \ref{tau_d_exvar} exhibits the normalized excess variance \citep{vau03} at all the above wave bands. In the GeV band, the variance is dominated by the uncertainties in the data. Therefore, the excess variance is negligible and is not shown on the table. Instead of $\tau_{d}$, the exponential growth or decay timescale given by $\tau_d/ln2$ may also be used as a characteristic timescale of variability. That will change the result by a constant factor and in most of the cases the value will be consistent with $\tau_d$ within the uncertainties.

The value of $\tau_d$ is the shortest at the hard X-ray band, slightly higher at the soft X-ray band and much longer at the optical, NIR as well as GeV bands. This is consistent with the assumption that the X-rays are produced by the highest energy tail of the electron distribution via synchrotron process. A consequence of that assumption is that the variability at the GeV band, being produced by lower energy electrons via IC processes, will be less pronounced, which is also evident. 

We have found the minimum variability timescale to be 1.1 ks. Assuming the variability is dominated by radiation cooling due to synchrotron only, 
\begin{equation}
t_{cool} \simeq  7.7 \times 10^8 (1 + z) \delta^{-1} B^{-2} \gamma_{\rm eff}^{-1} s,
\end{equation}
where $z$, $\delta$, $B$, and $\gamma_{\rm eff}$ are the redshift of the source, Doppler factor of the jet, magnetic field in the emission region, and effective Lorentz factor of the emitting electrons, respectively \citep[e.g.,][]{zha19,ryb79}. 

The characteristic frequency of the electron distribution responsible for the emission at the synchrotron peak of the SED is given by \citep{ryb79}
\begin{equation}
\nu_{ch} = 4.2 \times 10^6 \frac{\delta}{1+z}\gamma_{\rm eff}^2 B ~Hz
\end{equation}
Using $z$= 0.031, $\delta = 20$, and $\nu_{ch} = 10^{18}$ Hz in the equations (2) and (3), we find $B=0.5$ Gauss and $\gamma_{\rm eff} = 1.6 \times 10^5$. These values are consistent within a factor of a few with those obtained from modeling the broad-band SED of Mrk 421 in the standard leptonic scenario \citep{abd11,ban19}. The value of $\delta$ and $\nu_{ch}$ tend to vary from one epoch to another, which may cause appropriate changes in the above estimates of $B$ and $\gamma_{\rm eff}$ by a factor of a few.

Alternatively, instead of assuming the value of $\delta$ and $\nu_{ch}$, we eliminate $\gamma_{\rm eff}$ from equations (2) and (3) and find
\begin{equation}
B^3 \delta \simeq 2.5 (1+z) (\nu_{ch}/10^{18})^{-1} \tau_d^{-2},
\end{equation}
where $\nu_{ch}$ is in Hz and $\tau_d$ is in ks. Hence, using the value of any one of the variables $\delta$, $B$, and $\nu_{ch}$ the other two may be constrained.

From equations (2) and (3) it can be estimated that the ratio of the effective Lorentz factors of electrons contributing to the emission at frequencies $\nu_i$ and $\nu_j$ is proportional to $(\nu_i/\nu_j)^{0.5}$.  Therefore, the ratio of synchrotron cooling timescales for emission at frequencies $\nu_i$ and $\nu_j$ are proportional to $(\nu_i/\nu_j)^{-0.5}$, which implies that the ratios of cooling timescales at soft X-rays, optical, and NIR to that in hard X-rays are $\sim$3, $\sim$70, and $\sim$110, respectively. The value of $\tau_d$ from Table \ref{tau_d_exvar} are consistent with the above within the uncertainties.

Let us suppose that the GeV emission in Mrk 421 is produced by the inverse-Compton scattering of the synchrotron photons generated in the jet itself by the same relativistic electrons (the so called ``synchrotron self-Compton or SSC process''). We assume that the up-scattered photons have a mean energy of 1 GeV and most of the seed photons are from the synchrotron peak energy for Mrk 421, which is $\sim$1 keV. Then using the simplified formula
\begin{equation}
\nu_f = \gamma_{\rm IC}^2\nu_i,
\end{equation}
where $\nu_i$ and $\nu_f$ are the frequency of photons before and after the IC up-scattering and $\gamma_{\rm IC}$ is the effective Lorentz factor of the energetic electrons, we can estimate that the electrons responsible for the GeV emission have Lorentz factor $\sim$10$^3$. Hence, GeV emission variability timescale will be similar to that at the optical-NIR wavelengths, which is consistent with our findings.

Alternatively, the energy of electrons responsible for the GeV emission may be obtained using the inverse-Compton cooling timescale given by \citep[e.g.,][]{chi99}
\begin{equation}
t_{cool}^{IC} \simeq  3.0 \times 10^7 U_{rad}^{-1} \gamma_{\rm IC}^{-1} s,
\end{equation}
where $U_{rad}$ is the radiation energy density. Using $t_{cool}^{IC} = 100$ ks and $U_{rad} = 10^{-2}$ erg\,cm$^{-3}$ we obtain $\gamma_{\rm IC} = 3 \times 10^4$. Here, we estimate $U_{rad}$ to be approximately equal to the magnetic energy density. The obtained value of $\gamma_{\rm IC}$ is uncertain due to the uncertainties in $U_{rad}$ and $t_{cool}$.
\begin{figure}[!t]
\includegraphics[width=.8\columnwidth]{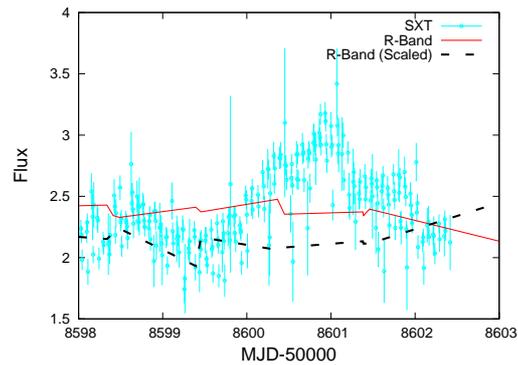}
\caption{The cyan open circles with error bars denote the SXT light curve of Mrk 421 during 2019 April 23-28. The red solid line corresponds to the $R$-band light curve during the same interval. The black dashed line indicates the $R$-band variability artificially generated based on the SXT light curve and assuming synchrotron origin of the emission at both wave bands. The properties of the actual and artificial optical light curves are consistent with each other. }\label{op_sim}
\end{figure}

The above estimates indicate that the optical variability should be similar but slower than the X-ray variability by a factor of $\sim$70 if both are produced by synchrotron radiation by different parts of the electron energy spectrum. To test this we scale the X-ray light curve appropriately to generate an artificial optical light curve, which is correlated with the X-ray variability but is slower by the above factor. We include some additional random fluctuations generated from the uncertainty of the SXT data. Figure \ref{op_sim} shows the artificial and the actual R-band light curves along with the SXT light curve and it is evident that the nature of the the real and artificial optical light curve segments are consistent with each other. 

\section{Discussion}
It can be seen from Table \ref{tau_d_exvar} that in the above observations, the variability timescale is shortest at the hard X-ray band and increasingly longer at soft X-ray, optical and NIR bands as well as in the GeV band. This is consistent with the model in which the X-ray to NIR (and longer wavelength) emission in Mrk 421 is generated by synchrotron radiation by the relativistic electrons in the jet. We find that the GeV emission may be produced via SSC process by electrons having energy lower than those producing the X-rays. Hence, the slower variability timescale of GeV emission is also justified. However, the energy density of the seed photons which are up-scattered to produce the GeV emission is not well-constrained in this case. Therefore, the estimate of the electron energy contains the corresponding uncertainty.  The obtained value of the magnetic field is consistent with those obtained from studies involving fitting the SED with physical models \citep[e.g.,][]{abd11}. 

The amplitude of variability, as indicated by the normalized excess variance, is also larger at the X-ray than the other wave bands. If the variability is caused by injection of energy or fluctuation in the magnetic field the amplitude of variability may not be different at the above wave bands. However, if the variability is faster in the X-ray band, it is obvious that in a given interval the observed amplitude will be larger. We note that while the amplitude of variability we find here is similar to that found by \citet{zha19} for Mrk 421 from Suzaku data the shortest variability timescale obtained here is an order of magnitude smaller than that found by \citet{zha19}. This may be due to the difference in sampling among the X-ray light curves or due to a different brightness state of Mrk 421.

It can be seen from Table \ref{tau_d_exvar} that $\tau_d$ during increase and decrease at all wavebands are not systematically different. In blazar variability it is usually assumed that the timescale of acceleration of electrons to the highest energies (e.g., $\gamma_{\rm eff}=10^5$ in this case) is faster than the cooling timescale such that the acceleration may be considered instantaneous. However, that will indicate $\tau_d$ during increase will be systematically shorter than that during decrease, which is not the case here. This implies that the acceleration timescale may not be negligible compared to the cooling timescales and may be constrained using short-timescale variability as analyzed here.

We note that the variability of the polarization fraction and electric vector polarization angle (EVPA) is more pronounced than that of the optical flux. This may be due the effect of turbulence \citep{mar14, lio20}. More data are needed to draw detailed inference about the polarization variability. 

Blazars vary at a large range of timescales often including minutes to years. The above quantitative results, obtained from the shortest end of the large range of timescales at which blazars vary, are consistent with the standard leptonic scenario, in which the lower energy peak of the spectral energy distribution of Mrk 421 is due to synchrotron radiation by the relativistic electrons in the jet and is located at the X-ray frequencies. Similar observation of a sample of blazars by AstroSat along with a more dedicated simultaneous multi-wavelength coverage will provide more specific constraints on the physical properties of the blazar class.


\section*{Acknowledgements}
\vspace{-1em}
We thank the anonymous referee for comments and suggestions that made the manuscript more comprehensive. RC thanks Presidency University for support under the Faculty Research and Professional Development (FRPDF) Grant, ISRO for support under the AstroSat archival data utilization program, and IUCAA for their hospitality and usage of their facilities during his stay at different times as part of the university associateship program. RC received support from the UGC start-up grant. RC acknowledges financial support from BRNS through a project grant (sanction no: 57/14/10/2019-BRNS) and thanks the project coordinator Pratik Majumdar for support regarding the BRNS project. This work has made use of data from the AstroSat mission of the ISRO, archived at the Indian Space Science Data Centre (ISSDC). This work has been performed utilizing the calibration databases and auxiliary analysis tools developed, maintained and distributed by AstroSat-SXT and AstroSat-LAXPC teams with members from various institutions in India and abroad and the SXT and LAXPC Payload Operation Center (POC) at the TIFR, Mumbai for the pipeline reduction. We are also thankful to the AstroSat Science Support Cell hosted by IUCAA and TIFR for providing the necessary data analysis software. The work has made use of software, and/or web tools obtained from NASA's High Energy Astrophysics Science Archive Research Center (HEASARC), a service of the Goddard Space Flight Center and the Smithsonian Astrophysical Observatory.

\balance



\end{document}